\documentclass[article,nojss]{jss}
\usepackage{amsmath,amssymb,booktabs,longtable,array}
\shortcites{Pedregosa2011}
\newcommand{\ComparativeReplicates}{24000}
\newcommand{\ComparativeRecords}{96000}
\newcommand{\LoadMedian}{0.01}
\newcommand{\LoadMinimum}{0.00}
\newcommand{\LoadMaximum}{0.03}
\newcommand{\MillionMetric}{0.39}
\newcommand{\MillionGaussian}{0.04}
\newcommand{\MillionPoisson}{0.15}
\newcommand{\MillionMultinomial}{0.17}
\newcommand{\LargeFit}{67.89}
\newcommand{\LargeBake}{11.59}
\newcommand{\LargePeak}{13.12}
\newcommand{\ParityError}{$5.55 \times 10^{-16}$}
\newcommand{\CoveragePercent}{94.6}
\newcommand{\StressSuccessful}{600}

\newcommand{\MetricTestCount}{383}
\newcommand{\MetricCoverage}{93.82}
\newcommand{\MetricLoadMedian}{0.00}
\newcommand{\MetricLoadMin}{0.00}
\newcommand{\MetricLoadMax}{0.02}
\newcommand{\MetricCoverageLow}{89.2}
\newcommand{\MetricCoverageHigh}{96.3}
\newcommand{\MetricParity}{$8.7\times10^{-13}$}
\newcommand{\ConformalExchangeable}{90.09}
\newcommand{\ConformalRatio}{91.43}
\newcommand{\ConformalArbitrary}{79.47}

\newcommand{\AttritionTests}{442}
\newcommand{\AttritionCoverage}{93.19}

\author{Sunday A. Adetunji\\Oregon State University
  \And Oluwatimileyin A. Adetunji\\Bowen University}
\Plainauthor{Sunday A. Adetunji, Oluwatimileyin A. Adetunji}
\title{\pkg{svylearn}: Design-consistent statistical learning, calibration, and inference for complex surveys}
\Plaintitle{svylearn: Design-consistent statistical learning, calibration, and inference for complex surveys}
\Shorttitle{Design-consistent statistical learning with svylearn}
\Abstract{Standard machine-learning workflows assume independent, identically distributed observations, leading to optimistic risk estimates, miscalibrated probabilities, and invalid uncertainty quantification when applied to complex probability samples. We introduce \pkg{svylearn}, an \proglang{R} package for design-consistent statistical learning, recalibration, and inference across cross-sectional and longitudinal surveys. Built on a strict separation between model development and evaluation, \pkg{svylearn} partitions data by whole primary sampling units (preserving all waves of longitudinal subjects), learns weighted preprocessing and hyperparameter tuning strictly inside training folds, and applies nested out-of-fold recalibration with auditable stabilization rules for rare outcomes. A unified evaluation interface (\code{svy_eval()}) reports design-based proper scores, rank metrics, calibration diagnostics, clinical utility, covariate-shift transport checks, weighted split-conformal intervals, and bounded-weight sensitivity analyses, explicitly distinguishing conditional fixed-score inference from complete-workflow replication. The package uses a lightweight S3 architecture with zero required third-party modeling dependencies at attachment, delegating estimation to optional learners (\pkg{glmnet}, \pkg{ranger}, \pkg{xgboost}, \pkg{nnet}) and survey variance calculations to \pkg{survey}. We verify numerical parity against \pkg{survey} and independent \proglang{Python} reference kernels, evaluate finite-sample behavior across extensive Monte Carlo simulations and stress tests, and demonstrate applied workflows using NHANES and longitudinal ECLS-K:2011 data. The baseline binary evaluation kernel processes one million records in 0.39 seconds, providing survey analysts with an auditable, reproducible end-to-end prediction pipeline.
}
\Keywords{complex surveys, cross-validation, calibration, rare outcomes, panel data, reproducibility, \proglang{R}}
\Plainkeywords{complex surveys, cross-validation, calibration, rare outcomes, panel data, reproducibility, R}
\Address{Sunday A. Adetunji\\Oregon State University\\
E-mail: \email{adetunjs@oregonstate.edu}\\
URL: \url{https://setrojalabs.github.io/}\\[1ex]
Oluwatimileyin A. Adetunji\\Bowen University\\
E-mail: \email{eniola16.to@gmail.com}}
\begin{document}

\section{Introduction}\label{sec:intro}

Survey analysts increasingly use flexible prediction models to summarize
heterogeneity in health, economic and social outcomes. A model trained on a
probability sample is embedded in two systems: the learning procedure that
maps observations to predictions, and the sampling design that determines
which observations are available. Validating only the learning algorithm can
leave the population target ambiguous. In particular, unequal selection
probabilities affect both fitting and performance evaluation, while clustered
sampling affects which observations can be treated as independent validation
units. Survey-aware cross-validation explicitly recognizes these distinctions
\citep{Wieczorek2022}. Probability calibration adds a separate requirement:
useful ranking does not establish agreement between predicted risks and
event frequencies \citep{VanCalster2019}.

Our starting point is the method of \citet{Adetunji2026}, which combined
whole-primary-sampling-unit (PSU) validation, weighted proper scores and
cross-fitted recalibration for rare cardiovascular-kidney-metabolic outcomes.
That study supplied the clinical motivating example and a survey-plasmode
benchmark. It reported an average nominal 95\% log-loss interval coverage of
30.2\% for naive validation and 95.0\% for a weighted, design-aware calibrated
strategy. These are empirical results of that experiment. They are not
universal coverage probabilities and are not newly generated software results.
Section~\ref{sec:validation} independently reconstructs the reported summaries
from the supplied replicate-level files before introducing new experiments.
We retain the software title cited in that article's Figshare reference
\citep{AdetunjiSoftware2026}. The NHANES application motivates the work;
the software accepts user-defined survey variables across subject domains
and supports both cross-sectional and longitudinal panel designs.

The contribution of \pkg{svylearn} is a reusable software implementation with
stronger separation of training and evaluation operations, a broader design
interface and an independently testable numerical core. It stores an explicit
sampling specification, allocates entire clusters to folds, learns every
preprocessing transformation inside a training set, selects candidate models
on inner out-of-fold scores, and trains the calibrator within the same outer
training set. It also records sparse-data fallbacks and distinguishes
conditional fixed-score uncertainty from complete workflow replication.
These engineering decisions address concrete sources of ambiguity when a
study-specific analysis is converted to a general package.

The package builds on established software. \pkg{survey} supplies sampling
design representations, linearization, replication and calibration of sampling
weights \citep{Lumley2004,Lumley2010}. \pkg{glmnet} supplies penalized generalized
linear models \citep{Friedman2010}; \pkg{ranger} supplies random forests
\citep{Wright2017}; \pkg{xgboost} supplies gradient-boosted trees
\citep{XGBoost}; and \pkg{nnet} supplies a feedforward neural-network adapter
\citep{Venables2002}. The fit/predict interface follows a familiar separation
of specification and estimation, also used in general machine-learning
software such as \pkg{scikit-learn} \citep{Pedregosa2011}.
Survey-aware splitting itself is established methodology, and the
\pkg{surveyCV} implementation already supports custom design-based validation
loops \citep{Wieczorek2022}. Our integration emphasizes nested calibration,
auditable sparse-data behavior and consistent reporting across these steps.

\section{Statistical targets and workflow}\label{sec:methods}

\subsection{Population risk and survey weights}

Let $U=\{1,\ldots,N\}$ denote a finite population, $s\subset U$ a probability
sample, $\pi_i>0$ the inclusion probability of unit $i$, and $w_i=1/\pi_i$ its
sampling weight. With binary outcome $y_i$ and probability prediction $p_i$,
the finite-population loss of a fixed prediction function is
\begin{equation}\label{eq:risk}
 R_U(p)=N^{-1}\sum_{i\in U}\ell(y_i,p_i).
\end{equation}
If $N$ is known, the Horvitz--Thompson risk estimator divides the weighted
sample loss total by $N$. The default in \pkg{svylearn} is the H\'ajek ratio
\begin{equation}\label{eq:hajek}
 \widehat R_H(p)=\frac{\sum_{i\in s}w_i\ell(y_i,p_i)}{\sum_{i\in s}w_i}.
\end{equation}
Multiplying all weights by a positive constant therefore leaves the reported
point metrics unchanged. We use logarithmic loss
$\ell_{\log}(y,p)=-y\log p-(1-y)\log(1-p)$ and the Brier loss
$\ell_B(y,p)=(y-p)^2$. Probabilities are clipped to $[10^{-15},1-10^{-15}]$
only where logarithms or logits require it; the Brier score uses the supplied
probabilities. This numerical convention is exposed in the documentation and
tested at the endpoints.

The distinction between a fixed $p$ and a prediction function learned from
$s$ is essential. For a learning rule $A$, $R_U(A(s))$ varies with the training
sample. Cross-validation estimates performance using smaller training sets
and can differ from the risk of the final full-sample fit even without
information leakage. The comparative experiment measures this discrepancy
directly. It does not relabel every observed difference as preventable bias.
Its census risk includes sampled units as members of the finite population;
it is not the risk on a disjoint external test population. This distinction
also contributes to the discrepancy from cross-validation.

For panel data, the unit in Equation~\ref{eq:risk} must be specified. A
person-wave target averages across eligible person-wave records and generally
requires wave-specific selection and response probabilities. A person-level
trajectory target is a different estimand. The software accepts user-supplied
longitudinal weights and checks nesting. It can also estimate sequential
retention weights inside nested validation from a baseline roster; the
identifying nonresponse assumptions are stated separately.

\subsection{Sampling geometry and fold construction}

The design object stores ordered sampling-stage identifiers, strata, optional
finite population corrections (FPCs), and optional replicate weights.
Inference uses these declarations through \pkg{survey}. For validation,
the highest-stage PSU is the indivisible unit. The implementation forms
collision-resistant keys from stratum and PSU identifiers and rejects a
subject or household that crosses the declared highest-stage clusters.
In a panel without a higher cluster identifier, the subject is the partition
unit. A duplicate subject-wave record is rejected.

Folds are allocated within strata while balancing the number of clusters and
the total sampling weight assigned to each fold. Randomness is controlled by
a supplied seed, and the caller's random-number state is restored. Outcomes
do not enter the allocation rule. This keeps partition construction separate
from outcomes and makes the outer independence test particularly direct.
With fewer than $K$ PSUs in a stratum, representation of that stratum in all
$K$ validation folds is impossible; the allocator distributes the available
PSUs and reports the design support. Zero-event validation folds remain in
the analysis. Their proper scores are defined, while single-class rank
metrics are returned as missing.

\subsection{Nested fitting and calibration}

For each outer validation set $V_k$, the pipeline operates only on
$T_k=s\setminus V_k$. Within $T_k$, an inner whole-cluster partition generates
out-of-fold predictions for each learner candidate. Each inner training
sample independently estimates weighted numeric medians, categorical modes,
scaling parameters and optional winsorization bounds or natural-spline
bases. Categorical encoding uses training levels; unseen validation levels
map to the training mode. The same frozen transformations are applied to
the corresponding validation records.

Candidate selection minimizes the inner weighted logarithmic loss. An
optional convex ensemble estimates nonnegative weights summing to one from
the inner prediction matrix. Denoting the selected or combined inner
predictions by $q_i$, a weighted sigmoid calibrator estimates
\begin{equation}\label{eq:platt}
 \Pr(Y_i=1\mid q_i)=\operatorname{logit}^{-1}
       \{a+b\operatorname{logit}(q_i)\},\qquad i\in T_k.
\end{equation}
Weighted isotonic and natural-spline calibration maps are also available.
The selected base learners are refitted on $T_k$, their probabilities are
transformed by this training-only calibration map, and the resulting
predictions are stored for $V_k$. The final model repeats the same training
procedure using all available observations, with inner cross-fitting to
construct its calibrator.

This ordering provides an implementation invariant: conditional on a fixed
design and random seed, changing outcomes in $V_k$ cannot change the pipeline
used to predict $V_k$. Every fitted operation is a function of $T_k$ alone.
An automated intervention test changes all outcomes in one outer validation
fold and verifies equality of that fold's predictions to numerical tolerance.
Each training pipeline has its own restored random-number stream, so an
earlier fold's outcome-dependent random-number consumption cannot influence
a later held-out fold. A stochastic-callback intervention test exercises
this dependency explicitly.
This checks the actual computational dependency, rather than relying only
on the name ``cross-fitting.'' Fitting a new calibrator to an arbitrary
collection of existing out-of-fold scores does not, by itself, establish this
invariant: the models that generated calibration-training scores may already
have used the outer validation outcomes.

Nested calibration can still experience a distributional change between
inner-fold predictions and predictions from a base learner refitted on more
observations. The outer validation loop evaluates this whole procedure,
including that change. It does not assume perfect calibration after fitting.

\subsection{Rare-event behavior and explicit stabilization}

Some numerical engines cannot fit a binomial model when a training class has
fewer than two observations. The package detects that condition before
calling the engine. It uses a smoothed intercept probability
\begin{equation}\label{eq:sparse}
 \widetilde p=\frac{n_{\mathrm{eff}}\widehat\pi+1/2}{n_{\mathrm{eff}}+1},\quad
 n_{\mathrm{eff}}=\frac{(\sum_iw_i)^2}{\sum_iw_i^2},\quad
 \widehat\pi=\frac{\sum_iw_iy_i}{\sum_iw_i}.
\end{equation}
This is a transparent stabilization rule for prediction, not an assertion
of Firth bias reduction or a design-based posterior. The effective sample
size in Equation~\ref{eq:sparse} measures weight concentration and does not
replace the survey design degrees of freedom.

For calibration, both classes must contain at least five raw and five Kish
effective observations before a slope is estimated by default. Otherwise,
the software holds the slope at one and estimates an intercept shift so
the weighted mean recalibrated probability equals $\widetilde p$.
Constant-score or failed calibration fits use a recorded constant fallback.
These rules avoid undefined predictions and make the price of limited event
information visible in the audit. Their performance is evaluated, not assumed.

\subsection{Continuous, count and multiclass responses}

The same population-risk target and nested partition contract apply beyond
binary classification. The \code{family} argument selects the response
representation and the inner selection loss. Continuous responses use
$\ell(y,\mu)=(y-\mu)^2$. Nonnegative counts use Poisson deviance,
\begin{equation}
 \ell_D(y,\mu)=2\{y\log(y/\mu)-y+\mu\},
\end{equation}
with $0\log(0/\mu)=0$ and a numerical floor $10^{-15}$ for $\mu$.
This loss evaluates a conditional mean; choosing it does not establish that
survey counts have Poisson variance. Multiclass responses use
$-\log p_y$, with factor levels declaring the class schema before splitting.
The multiclass Brier score is $\sum_k(\mathbf{1}\{y=k\}-p_k)^2$, without
division by the number of classes. Classification ties use the first declared
class for the reported accuracy. All families allow categorical and numeric
predictors and the same training-fold missing-predictor recipe.

Automatic recalibration is fitted only to inner out-of-fold predictions.
For continuous responses it estimates a weighted affine map $a+b\mu$;
for counts it multiplies the mean by $\sum_iw_i y_i/\sum_iw_i\mu_i$.
Degenerate inputs use an explicit weighted-mean fallback. Multiclass
recalibration uses $q_k=p_k^{1/T}/\sum_jp_j^{1/T}$, with $\log T$ optimized
over $[-4,4]$ by weighted log loss. This adapts temperature scaling
\citep{Guo2017} to the weighted inner sample. An optimum at the search
boundary is recorded. Temperature scaling preserves class ordering, but
none of these estimated recalibrations is guaranteed to improve unseen risk.

The default multiclass learner is a baseline-category softmax GLM with a
$10^{-6}$ slope ridge penalty, excluding intercepts, applied to mean weighted
log loss. Its analytic gradient is supplied to BFGS optimization. The
\pkg{glmnet}, \pkg{ranger} and \pkg{xgboost} adapters also cover all four
families. The \pkg{nnet} adapter supports binary, continuous and multiclass
responses. Count forests estimate means through regression and are evaluated
by Poisson deviance. When any declared class has fewer than two training
observations, a recorded effective-sample-size smoothed class-frequency model
replaces that fit. Declared but absent classes retain positive probabilities.
The new mean-loss inference, permutation summaries and complete replicate
refits use the same family-specific scores. Binary discrimination and decision
curves retain their binary definitions.

\subsection{Discrimination, calibration and decision summaries}

Weighted ROC area is computed as the weighted probability that a case has a
higher score than a noncase, with half credit for ties. Aggregation at
distinct score thresholds avoids a quadratic case-by-control comparison.
Average precision integrates a stepwise precision-recall curve and handles
all observations with the same score together. It is not a trapezoidal PR
area. Partial ROC area integrates linearly interpolated ROC segments up to
the requested false-positive rate and is returned without standardization.

Calibration output includes an offset-model intercept, a joint-model slope,
and their survey-based standard errors. Optional ICI, E50 and E90 summarize
the weighted absolute discrepancy between a natural-spline calibration
smoother and the supplied probabilities. These diagnostics depend on the
smoothing choice. Equal-width calibration bins are provided separately and
are not labeled as ICI. Decision summaries include weighted confusion totals,
sensitivity, specificity, positive and negative predictive values, and net
benefit $\widehat{TP}/\widehat N-
\{t/(1-t)\}\widehat{FP}/\widehat N$ at threshold $t$.

\subsection{What the uncertainty estimates cover}

For fixed predictions and a H\'ajek mean, the linearized contribution is
$e_i=w_i\{\ell_i-\widehat R_H\}/\sum_jw_j$. In a first-stage stratified
design with $m_h$ sampled PSUs and no FPC, the ultimate-cluster variance is
\begin{equation}\label{eq:variance}
 \widehat V=\sum_h\frac{m_h}{m_h-1}
   \sum_{c=1}^{m_h}(E_{hc}-\overline E_h)^2,\qquad
 E_{hc}=\sum_{i\in(h,c)}e_i.
\end{equation}
The package delegates multistage and FPC calculations to \pkg{survey}; it
does not substitute Equation~\ref{eq:variance} for all designs. Proper-score
intervals use a Student reference distribution with survey design degrees
of freedom. Domain evaluation subsets the parent design. Singleton policies
are explicit, and global survey options are restored after evaluation.

For a supplied replication scheme, the variance is formed from the scheme's
scale, replicate-specific factors and centering convention. Bootstrap, BRR,
Fay, JK1 and JKn therefore retain their own conventions. Rank-metric
replication recomputes the metric at each replicate weight vector while
holding predictions fixed. A separate interface refits preprocessing,
selection and calibration at every replicate, thereby replicating the
complete nested statistic. Failed refits remain in the output and prevent
a variance estimate; they are not silently discarded.

Neither procedure implies exact coverage of the risk of every learned
model. Fixed-score linearization omits training variation and dependence
created by overlapping training sets. The difficulty of universal variance
estimation for cross-validation is established even in simpler sampling
settings \citep{Bengio2004}. Complete replication also requires
an appropriate design and regularity of the statistic. We expose the
uncertainty target in the returned object and assess these distinctions
empirically in Section~\ref{sec:validation}.

\subsection{A fixed evaluation contract}\label{sec:metricsuite}

Version 0.2.0 adds a common reporting interface, \code{svy_eval()}, with
12 core diagnostic families and 17 glossary rows. Each key has a specified
estimand and an availability condition. For example, transport balance
requires a target sample, while a post-raking event residual requires a
known event total. A missing input produces an explicit unavailable result;
it does not produce a numerical zero. We do not claim that these constituent
estimators are newly invented. The contribution is their explicit integration
with survey-aware prediction, reproducible computation and validation.

Let $a_i=w_i/\sum_jw_j$, $\ell_i=\ell(Y_i,\widehat p_i)$, and
$\widehat L=\sum_i a_i\ell_i$. The \code{optimism_bias} key preserves the
specified contrast
\begin{equation}
 \widehat\Delta=\frac1n\sum_i\ell(Y_i,\widehat p_i^{\mathrm{IID}})
              -\sum_i a_i\ell(Y_i,\widehat p_i^{\mathrm{survey}}).
\end{equation}
Both prediction vectors must be genuinely out of sample. This observed
contrast combines changes in fitting, weighting, splitting and calibration;
it is not an identified decomposition of true optimism. Its sign is unrestricted.
For \code{deff_loss} and \code{n_eff_pred}, the frozen reference is
\begin{equation}
 \widehat{\mathrm{DEFF}}_\ell=
 \frac{\widehat{\mathrm{Var}}_d(\widehat L)}{s_\ell^2/n},
 \qquad \widehat n_{\mathrm{eff}}=n/\widehat{\mathrm{DEFF}}_\ell,
 \qquad s_\ell^2=\frac{\sum_i(\ell_i-\bar\ell)^2}{n-1}.
\end{equation}
This unweighted, with-replacement SRS denominator is deliberately distinguished
from other survey design-effect conventions. A zero reference variance makes
the ratio undefined. Effective sample size can exceed $n$ under a variance-reducing
design; it is a variance equivalence, not a theorem about model complexity or power.

For binary outcomes, the spline diagnostic $\widehat g_w$ gives
\begin{equation}
 e_i=|\widehat p_i-\widehat g_w(\widehat p_i)|,\qquad
 \widehat{\mathrm{ICI}}=\sum_i a_i e_i,\qquad
 \widehat E_{90}=\inf\{e:\sum_i a_i\mathbf{1}(e_i\le e)\ge .9\}.
\end{equation}
These adapt the calibration summaries of \citet{Austin2019}. The implementation
uses a weighted logistic natural-spline smoother with a documented sparse-event
fallback. Replicate evaluation refits this smoother, holding the prediction rule
fixed. The offset intercept (calibration in the large) is separate from the free
intercept in a joint intercept/slope regression. For raked weights $w_i^r$,
\begin{equation}
 \widehat R_{\mathrm{rake}}=
 \frac{\sum_iw_i^r\widehat p_i-T_Y}{T_Y},\qquad T_Y>0.
\end{equation}
Covariate calibration does not force this event residual to zero. The function
also reports covariate-margin residuals separately, including absolute residuals
when a target margin is zero.

Transport diagnostics compare source and target covariate means using a fixed
reference scale,
\begin{equation}
 \mathrm{SMD}_j=\frac{\bar X_{S,j}-\bar X_{T,j}}
 {\{(v_{S,j}+v_{T,j})/2\}^{1/2}},\qquad
 \widehat D=\sum_{i\in T}a_{T,i}
 (\widehat p_{S\rightarrow T,i}-\widehat p_{T,i})^2.
\end{equation}
Here $v_{S,j}$ and $v_{T,j}$ are baseline weighted population-moment variances;
the denominator remains fixed after applying supplied transport weights.
Both prediction vectors in $\widehat D$ are evaluated on the same target rows.
Target outcome Brier loss is a different output. Neither agreement nor small
SMD proves transportability, positivity or causal exchangeability.

\subsection{Prediction sets and bounded-weight sensitivity}

Weighted split conformal uses calibration scores $V_i$ from an independently
fitted prediction rule. If $r_i$ and $r(x)$ are appropriate density ratios, its
threshold is the $(1-\alpha)$ quantile of
\begin{equation}
 \sum_{i=1}^m\frac{r_i}{\sum_jr_j+r(x)}\delta_{V_i}
       +\frac{r(x)}{\sum_jr_j+r(x)}\delta_{+\infty}.
\end{equation}
The infinity atom is retained, so insufficient calibration mass can produce
an infinite regression interval. Equal weights recover the exchangeable split
construction. The covariate-shift interpretation requires the conditions in
\citet{Tibshirani2019}; survey inclusion weights alone do not establish them.
On a separate test probability sample, the reporting keys are empirical summaries
\begin{equation}
 \widehat C=\sum_{i\in T}a_i\mathbf{1}\{Y_i\in\Gamma(X_i)\},\qquad
 \widehat W=\sum_{i\in T}a_i|\Gamma(X_i)|,
\end{equation}
where width means interval length for regression and cardinality for classification.
These outputs do not claim distribution-free conditional coverage, which cannot
hold generally without further restrictions \citep{Barber2021}.

For finite row-level utilities $b_i$, \code{svy_weight_bounds()} returns the sharp
point bounds over $u_i\in[\Gamma^{-1},\Gamma]$:
\begin{equation}
 B_-(\Gamma)=\min_u\frac{\sum_iw_i u_i b_i}{\sum_iw_i u_i},\qquad
 B_+(\Gamma)=\max_u\frac{\sum_iw_i u_i b_i}{\sum_iw_i u_i}.
\end{equation}
Sorting $b_i$ reduces the lower-bound optimization to $n+1$ threshold tilts:
values below the minimizing mean receive the upper multiplier. Thus no stochastic
optimizer is needed. The default \code{sensitivity_gamma} is the first $\Gamma$
at which $B_-(\Gamma)$ reaches a specified null utility, located within a reported
search interval. An optional \code{tilted_lower_ci} target recomputes a conditional
survey ratio interval along the mean-minimizing path. It reports the first
grid-bracketed crossing, and explicitly does not claim to minimize the lower
confidence bound globally. Such a confidence-bound optimization is different
from the linear-fractional point problem. Neither target identifies the magnitude
of unmeasured confounding or a missing-outcome mechanism without additional assumptions.

\subsection{Utility, domain calibration and explanations}

At threshold $t$, define $I_i(t)=\mathbf{1}(\widehat p_i\ge t)$ and weighted
confusion totals $T_{\mathrm{TP}}=\sum_iw_iY_iI_i$,
$T_{\mathrm{FP}}=\sum_iw_i(1-Y_i)I_i$, with analogous TN and FN totals.
The utility and screening outputs are
\begin{equation}
 \widehat{\mathrm{NB}}(t)=\frac{T_{\mathrm{TP}}-tT_{\mathrm{FP}}/(1-t)}{\sum_iw_i},
 \quad N_{\mathrm{follow}}=\frac{T_{\mathrm{TP}}+T_{\mathrm{FP}}}{T_{\mathrm{TP}}},
 \quad N_{\mathrm{screen}}=\frac{\sum_iw_i}{T_{\mathrm{TP}}}.
\end{equation}
The \code{nns_svy} key retains $N_{\mathrm{follow}}$, the positive follow-ups
per true positive. The second key, \code{nns_population}, gives
$N_{\mathrm{screen}}$ when the whole population is screened.
These denominators must not be interchanged. Counts
require population-scale weights, or a supplied population size multiplying
estimated proportions. Person-wave totals are not automatically counts of
distinct people. Utility is also reported per 100,000, with treat-all and
treat-none comparators. Intervals are conditional and pointwise across thresholds.

A joint survey-weighted domain calibration model estimates
$\operatorname{logit}\Pr(Y_i=1)=\alpha_d+\beta_d\operatorname{logit}(\widehat p_i)$.
The \code{domain_cal_gap} output is
\begin{equation}
 G_\alpha=\max_d|\widehat\alpha_d|,\qquad
 G_\beta=\max_d|\widehat\beta_d-1|.
\end{equation}
Unsupported domains are identified before fitting. Pairwise contrasts use the
full joint coefficient covariance, including covariance between domains sharing
sampling units. Conservative maximum-gap envelopes are obtained from a Bonferroni
coefficient rectangle. These are descriptive calibration departures rather than
a complete fairness or policy-effect assessment.

Finally, supplied local SHAP values $\phi_{ij}$ and within-PSU perturbations
$X_i^{(b,j)}$ give
\begin{equation}
 \widehat\Phi_j=\sum_i a_i|\phi_{ij}|,\qquad
 \widehat{\mathrm{VIP}}_j=\frac1B\sum_b\sum_i a_i
 \{\ell(Y_i,\widehat f^{(-k(i))}(X_i^{(b,j)}))-\ell_i\}.
\end{equation}
The SHAP background and explanation algorithm remain external inputs.
For panel data, permutation moves whole subject trajectories among subjects
with compatible wave schedules within the same sampling unit. Conditional
design intervals treat fitted rules and realized permutations as fixed;
Monte Carlo permutation variability is reported separately. These quantities
describe model dependence and are not causal feature effects.

\section{Software architecture}\label{sec:software}

\pkg{svylearn} uses small S3 objects for design specifications, recipes,
learner specifications, calibrators and fitted workflows. A design object
stores the analysis frame and design metadata; it does not load a model
engine. A recipe stores training summaries and basis definitions. A learner
specification stores an engine name and parameters; fitting triggers the
optional namespace lookup. A fitted workflow retains outer predictions,
fold assignments, candidate losses and mixture weights, stabilization
statuses, final models and design diagnostics.

The implementation imports only packages shipped with \proglang{R}
\citep{R}. Survey inference and third-party learner packages are optional
dependencies loaded at the point of use. Package attachment performs no
network access, package installation, worker startup, input-data loading or
global-option modification. It uses no custom compiled code; computationally
intensive model fitting is delegated to established engines. This division
keeps installation portable and avoids paying an initialization cost for
engines that an analyst does not use.

Numeric preprocessing uses training weights, while candidate engines receive
weights normalized to mean one. The same rescaling is applied consistently
within each fit. It does not create a universal equivalence of penalty
definitions across engines. Hyperparameters retain the semantics of their
underlying software. Optional winsorization acts on predictor values and
does not silently trim sampling weights or change the declared population
target. Custom callbacks receive a numeric predictor matrix, an outcome vector
and training weights. Outcomes are numeric for scalar families and factors
with declared class levels for multinomial learning. Predictions are checked
for dimension, finiteness and family-specific support; multiclass probability
columns are aligned to the declared class schema. Probability sums within
$10^{-6}$ of one are normalized to remove floating-point roundoff.

Input validation rejects missing design variables, invalid probability
vectors, nonpositive fitting weights, malformed replicate matrices and
invalid panel nesting. Replicate evaluation permits zero weights because
replication schemes can omit sampled units. Zero-weight units are removed
before a complete refit. Missing predictors are handled through the
training recipe. Outcome learners reject missing outcomes; attrition rosters
retain explicitly unobserved rows for retention estimation.
An inner-fold engine failure excludes that candidate from selection and
stacking, with the failure retained in the candidate audit. If every
candidate fails, the workflow stops with the collected diagnostic rather
than returning an apparently successful fit. In particular, an incomplete
or nonconverged elastic-net path is not accepted as a valid candidate.

The supported capabilities and their numerical owners are summarized in
Table~\ref{tab:api}. Reusing \pkg{survey} for inference is a deliberate
choice: it provides a recognized reference implementation against which the
adapter can be tested independently.

\begin{table}[tbp]
\centering\small
\begin{tabular}{>{\raggedright\arraybackslash}p{.25\textwidth}p{.65\textwidth}}
\toprule
Interface & Responsibility \\\midrule
\code{svy_design} and \code{svy_split} & Validate design and panel identifiers; partition whole PSUs; audit weight and event support.\\
\code{svy_recipe} & Learn and freeze weighted imputation, scaling, winsorization, splines and optional thin-SVD PCA.\\
\code{svy_trajectory} & Summarize repeated measurements using a prespecified time cutoff.\\
\code{svy_fit} & Run nested validation, candidate selection, optional convex stacking and training-only calibration.\\
\code{svy_infer} & Delegate conditional score and calibration inference to \pkg{survey}; recompute rank metrics under replicate weights.\\
\code{svy_attrition_weights} & Estimate sequential retention weights; an unfitted specification triggers nested nuisance-model refitting.\\
\code{svy_refit_replicates} & Refit the complete workflow using supplied replicate-weight conventions.\\
\code{svy_subgroups} and \code{svy_shap} & Audit subgroup support and aggregate supplied model attributions with design-based uncertainty.\\
\code{svy_importance} & Measure out-of-fold loss changes after within-PSU predictor permutations.\\
\code{svy_eval} and \code{svy_glossary} & Report the frozen metric suite, estimands, availability and conditional uncertainty.\\
\code{svy_conformal} & Construct weighted split intervals or classification sets with explicit assumptions.\\
\code{svy_sensitivity} & Evaluate bounded-weight mean tipping and an explicitly distinguished conditional interval path.\\
\code{svy_export} & Write CSV, Markdown or LaTeX tables from evaluation objects.\\
\bottomrule
\end{tabular}
\caption{Public interfaces and their responsibilities. Attribution summaries are associational and depend on the supplied model and explanation background.}\label{tab:api}
\end{table}

\clearpage
\section{Illustrative workflows}\label{sec:workflows}

\subsection{Cross-sectional learning}

The following example uses a synthetic teaching dataset with ten strata and
four PSUs per stratum. It can be run without access to the motivating study.
The candidate list forms a small tuning grid for an elastic-net learner.

\begin{CodeInput}
R> library("svylearn")
R> d <- svy_example(1000, seed = 17)
R> design <- svy_design(d, "weight", "psu", "stratum")
R> learners <- list(
+    svy_learner("glmnet", list(alpha = 0.5, lambda = 0.005)),
+    svy_learner("glmnet", list(alpha = 0.5, lambda = 0.02)))
R> fit <- svy_fit(design, "y", c("x1", "x2"), learners,
+                 v = 5, inner_v = 3, repeats = 5, seed = 17)
R> summary(fit)
R> svy_export(fit, "performance.csv")
\end{CodeInput}

The summary separates mean repeat performance, repeat-to-repeat variation
and consensus-prediction performance. These quantities answer different
questions. The consensus vector averages probabilities across repeated
partitions and generally has a different loss from the mean loss of the
individual repeats. Repeats reuse the same sampled observations, so their
standard deviation is not an independent-sample standard error.

New observations are scored by \code{predict(fit, newdata)} using the
full-data final pipeline. Internal validation uses \code{fit\$predictions}
or \code{fit\$consensus}, not final-model predictions on its own training
data. The latter estimate apparent performance.

\subsection{Repeated observations and attrition weights}

The same API can classify person-wave outcomes while preserving subject
nesting. The synthetic example below has three waves per subject.

\begin{CodeInput}
R> panel <- svy_example(1000, waves = 3, seed = 31)
R> panel_design <- svy_design(panel, "weight", "psu", "stratum",
+                            subject = "subject", wave = "wave")
R> panel_fit <- svy_fit(panel_design, "y", c("x1", "x2", "wave"),
+                       learners, v = 5, inner_v = 3, seed = 31)
\end{CodeInput}

For an actual panel, \code{weight} must be replaced by the appropriate
analysis weight. Preserving subjects across folds protects an unseen-cluster
evaluation target; it does not by itself implement forecasting from earlier
to later waves. A temporal forecasting task requires a different validation
restriction and a corresponding target definition.

\subsection{Estimating retention weights inside nested validation}

For a full scheduled panel roster, let $R_{it}$ indicate observation at wave
$t$, with $R_{i0}=1$ and monotone dropout. In a training partition $A$, the
retention model estimates
\begin{equation}
 \widehat p_{it}^{A}=\widehat{\Pr}_{A}
 (R_{it}=1\mid R_{i,t-1}=1,H_{i,t-1},X_{i0},w_{i0}),\qquad
 \widehat w_{it}^{A}=w_{i0}\prod_{s=1}^{t}
 \frac{\widehat q_s^{A}}{\max(\epsilon,\widehat p_{is}^{A})}.
\end{equation}
Here $H_{i,t-1}$ is automatically lagged observed history, $X_{i0}$ contains
baseline predictors, and $w_{i0}$ is the baseline survey weight. The default
denominator model is survey-weighted logistic regression, fitted separately
in each wave's training risk set with normalized baseline weights. A single
prespecified \code{svy_learner} can replace the GLM. Retention preprocessing is
trained on that same risk set. By default $\widehat q_s^{A}=1$; optional
stabilization uses its weighted training retention fraction. Wave-specific
constant numerators preserve a normalized within-wave mean but change the
relative mass in a pooled person-wave estimand.

Define an unfitted specification with \code{svy_attrition()}.

The helper \code{svy_attrition_weights()} fits it on selected whole training
groups and predicts retention on the remaining roster. In the full
workflow, every inner training fold fits its own retention models before
weighted outcome preprocessing, fitting and validation. Inner out-of-fold
weights also enter tuning, stacking and recalibration. The outer-training
retention fit supplies weights for the outer evaluation fold. Repeat-specific
weights, risk-set counts, clipping counts and fallback statuses are retained.
The final fitted model uses all training data; internal evaluation continues
to use the separately retained out-of-fold predictions and weights.

For example, \code{panel} below includes dropout rows, a baseline weight
\code{w0}, a monotone observation indicator \code{observed}, and outcome
\code{y}. The prespecified logical column \code{target} selects eligible waves:
\begin{CodeChunk}
\begin{CodeInput}
R> roster_design <- svy_design(panel, "w0", "psu", "stratum",
+                             subject = "id", wave = "wave")
R> retention <- svy_attrition("observed", history = c("y", "x"),
+                            analysis = "target")
R> retention_fit <- svy_fit(roster_design, "y", "x", family = "gaussian",
+                          v = 5, inner_v = 3, attrition = retention)
R> summary(retention_fit)$attrition
\end{CodeInput}
\end{CodeChunk}

Complete scheduled rosters prevent absent records from silently disappearing
from a risk set; intermittent response and varying within-subject baseline
weights are rejected. Outcome learners use observed eligible rows only.
Sequential MAR, positivity and an adequate retention model are identifying
conditions, not consequences of software validation \citep{Metten2022}.
Probability clipping and sparse-class fallbacks are reported because they may
alter bias. The no-dropout case preserves baseline weights exactly.
For complete-workflow replication, supply replicate weights for the full
baseline roster; retention and outcome models are both refitted. Fixed-score
evaluation instead conditions on the estimated weights. The ECLS application
uses agency-provided longitudinal adjustments and does not add a second
estimated retention correction.

\subsection{Conditional inference and reporting}

The inference call takes predictions explicitly so that their provenance
can be inspected. The following calculations condition on the consensus
prediction vector. The bootstrap design can also be used to obtain
replicate-weight uncertainty for rank metrics.

\begin{CodeInput}
R> result <- svy_infer(design, d$y, fit$consensus)
R> result$scores
R> result$calibration
R> rep_design <- svy_replicates(design, type = "bootstrap",
+                              replicates = 500, seed = 17)
R> ranks <- svy_infer(design, d$y, fit$consensus,
+                    replicate_design = rep_design)$ranks
R> svy_export(ranks, "rank_metrics.tex", format = "latex")
\end{CodeInput}

Subgroup output includes raw sample and event counts and suppresses
calibration inference when minimum event support is not met. A caller can
inspect weighted discrimination or calibration curves, but uncertainty and
decision relevance still require an appropriate evaluation sample.
Similarly, supplying SHAP contributions allows a survey-weighted mean
absolute attribution and its conditional standard error; it does not make
the explanation causal or correct an inappropriate SHAP background.

\subsection{Other outcome families and subject domains}

An analyst supplies the outcome, predictor and sampling-variable names; no
NHANES naming convention is required. For example, the following uses a
continuous response in a clustered probability sample:
\begin{CodeChunk}
\begin{CodeInput}
R> study <- svy_example(500)
R> set.seed(42)
R> study$amount <- 2 + study$x1 + rnorm(nrow(study))
R> design_amount <- svy_design(study, "weight", "psu", "stratum")
R> fit_amount <- svy_fit(design_amount, "amount", c("x1", "x2"),
+    family = "gaussian", v = 3, inner_v = 2)
R> svy_infer(design_amount, study$amount, fit_amount$consensus,
+    family = "gaussian")$scores
\end{CodeInput}
\end{CodeChunk}
For count responses the family becomes \code{"poisson"}; for a factor
outcome it becomes \code{"multinomial"}. Multiclass predictions form an
observation-by-class-by-repeat array with named class columns. Scalar-response
predictions remain an observation-by-repeat matrix. Independent observations
use \code{svy_design(data, "weight")} without cluster identifiers, with
unit weights when appropriate. Declaring a panel adds the subject and wave
metadata shown above without changing the outcome-family interface.

\subsection{A fresh NHANES analysis}

The replication script reconstructs the published analytic cohort from all
18 original public-use NHANES XPT inputs, verifies their SHA-256 hashes,
and checks participant identifiers and outcomes against the supplied
prediction archive. It recovers 1,743 participants with 44 events and a
weighted prevalence of approximately 1.97\%. The outcome definition and
clinical interpretation remain those of \citet{Adetunji2026}.

We then fit a new 20-repeat, five-fold workflow with three inner folds.
The candidate library includes a logistic model, six elastic-net
specifications crossing $\alpha\in\{0,0.5,1\}$ with
$\lambda\in\{0.005,0.02\}$, a 200-tree random forest and a 100-round
boosted-tree model. Predictor construction is inherited from the published
analysis, while preprocessing and calibration are re-estimated within the
new nested workflow. The new numbers should therefore not be expected to
equal the earlier study's model comparisons.
The candidate audit excludes 12 nonconverged elastic-net evaluations among
900 outer-fold candidate evaluations and retains their diagnostics. The
remaining candidates permit every outer fold to complete.

Table~\ref{tab:nhanes} reports the new repeated-validation metrics and the
consensus metrics separately. Proper-score and rank-metric intervals retain
the parent MEC survey design when evaluating the analytic domain.
The software example concerns concurrent classification; it is not an
external validation of future CKM risk or a clinical deployment study.

\begin{table}[t]
\centering\small
\begin{tabular}{lrrrr}
\toprule
Metric & Mean repeat & Repeat SD & Consensus & Conditional 95\% interval \\
\midrule
Log loss & 0.0875 & 0.0025 & 0.0860 & [0.0647, 0.1074] \\
Brier score & 0.0191 & 0.0003 & 0.0190 & [0.0129, 0.0251] \\
ROC area & 0.7918 & 0.0301 & 0.8172 & [0.7737, 0.8608] \\
Average precision & 0.0710 & 0.0172 & 0.0722 & [0.0315, 0.1129] \\
\bottomrule
\end{tabular}
\caption{Fresh NHANES evaluation using 20 outer repeats and training-only nested calibration. Repeat SD describes partition instability; consensus intervals condition on the prediction vector and retain the parent survey domain design. Rank intervals use 500 survey-bootstrap replicates.}\label{tab:nhanes}
\end{table}

\subsection{Longitudinal education data: ECLS-K:2011}\label{sec:ecls}

To demonstrate use beyond a cross-sectional health survey, we use the official
public-use ECLS-K:2011 kindergarten--fifth-grade file \citep{NCES2019}.
The download and extraction scripts verify the archive, dictionary and extract
hashes and retain the original data separately. The file contains 18,174
children, of whom 8,542 have positive \code{W9C29P_9A0}. This longitudinal
weight adjusts for response to the spring child components from kindergarten
through fifth grade, the kindergarten parent component and the fifth-grade
parent component. It is selected from the manual's component-specific weights,
rather than replaced with a generic wave weight.

Reading and mathematics trajectories supply observed counts, means, last
measurements and slopes using data available no later than the preceding
grade. These features are constructed with \code{svy_trajectory()}; missing
predictor values are imputed inside training folds. Negative nonresponse code
$-9$ and blank IRT scores are missing outcomes, which are not imputed.
Age at kindergarten entry and recorded sex complete the predictor set; the
pooled panel model also includes target grade. The targets are released IRT
theta scores, not diagnoses or causal effects of educational interventions.

We evaluate fifth-grade reading and a pooled one-year prediction task over
grades 3--5. The latter contains repeated child-wave records, with a separate
time cutoff for each prediction. Five outer folds, three inner folds and five
repeats compare weighted least squares, ridge and elastic-net candidates;
inner out-of-fold predictions fit affine recalibration. Folds keep each released
variance unit, and hence every child's records, together. The original PSU
identifier is suppressed in the public file, so the partition claim refers
specifically to its 251 released variance units in 70 variance strata.

Inference returns predictions to the parent design and evaluates the
observed-outcome domain using the 80 corresponding JK2 replicate weights.
The variance is the sum of squared replicate deviations from the full estimate,
with the manual's 80 degrees of freedom. It conditions on the fitted consensus
predictions and does not refit the learning pipeline in each replicate.
The targets therefore concern observed-score longitudinal respondents and,
for the pooled task, their eligible child-wave records. No claim of automatic
correction for unobserved outcomes is made.

\begin{table}[tbp]
\centering
\begin{tabular}{lrrrr}
\toprule
Task & Records & Children & MSE & Conditional 95\% CI \\
\midrule
Fifth grade & 8,487 & 8,487 & 0.02742 & [0.02641, 0.02844] \\
Grades 3--5 panel & 25,453 & 8,497 & 0.02533 & [0.02478, 0.02587] \\
\bottomrule
\end{tabular}
\caption{ECLS-K:2011: consensus out-of-fold squared error on the released reading-theta scale. Five outer folds, three inner folds and five repeats are used. Intervals use the 80 official JK2 replicates and 80 degrees of freedom; pipeline refitting is not included. The panel target averages eligible child-wave records, not unique children.}\label{tab:ecls}
\end{table}

This application exercises a real longitudinal survey, time-eligible
preprocessing, person-level validation boundaries and agency-specific replication.
It complements NHANES by demonstrating an education outcome on a continuous
scale and makes the distinction between child totals and child-wave totals
explicit in the machine-readable design metadata.

\section{Validation experiments}\label{sec:validation}

We organize each simulation around its aim, data-generating mechanism,
estimand, competing methods and performance measures, following the ADEMP
reporting framework of \citet{Morris2019}. Fixed-score sampling checks the
inference adapter; comparative learning tests procedure behavior under
misspecification and attrition; development stress tests expose numerical
failure. These experiments have distinct targets and are reported separately.

\subsection{Numerical and software checks}

The verification suite compares weighted tied-score AUC with its exact
pairwise definition, checks invariance to weight rescaling, and tests
single-class and probability-boundary cases. Training-fold transformations
are serialized before and after validation to verify that applying a recipe
cannot alter it. Panel and PSU assertions check that every member of a
partition unit receives one fold assignment. The outcome-intervention test
described in Section~\ref{sec:methods} checks the nesting invariant directly.

Survey agreement tests cover multistage FPCs, domains, supplied replicate
weights, JKn, BRR, Fay and bootstrap conventions, singleton policies and
raking totals. Independent \proglang{Python} reference kernels reproduce
weighted score estimates and first-stage Taylor standard errors with a
maximum absolute discrepancy of \ParityError{}. This parity statement
applies to those kernels and that fixture; it is not a claim of identical
cross-language model engines. The baseline automated tests exercise \CoveragePercent\%
of instrumented source lines. The expanded version 0.2.0 suite has
\MetricTestCount{} passing expectations and \MetricCoverage\% measured line
coverage. This is a software coverage measure, separate
from confidence-interval coverage.

The additional response tests compare continuous and count fits with direct
weighted \proglang{R} GLMs and compare the native multinomial optimizer with
\code{nnet::multinom}. Poisson scores agree with the reference family's
deviance residual calculation; multiclass scores agree with direct indicator
matrix calculations. Weighted moment identities verify continuous and count
recalibration. Outer-label intervention tests are repeated for every new
family, including stochastic callbacks and stacking. Fourteen supported
engine--response combinations undergo prediction and fresh-process
serialization checks, including single-row multiclass predictions.

\subsection{Reaggregation of the originating benchmark}

The supplied archive contains 24,000 pipeline records: four strategies for
each of 6,000 simulated samples, arranged as 12 scenarios with 500 samples
each. We check unique scenario--replicate--pipeline keys, verify scenario
counts, and recompute bias, root mean squared error (RMSE), coverage and
Monte Carlo standard errors (MCSEs) from the replicate estimates and truths.
The recomputed summaries agree with the supplied summary table to better
than $10^{-12}$.

The calibrated strategy has average log-loss coverage 94.97\%, which rounds
to the published 95.0\%. Its scenario-specific values range from 91.4\%
to 97.4\%. The average absolute scenario bias is 0.00262. For naive random
validation, average coverage is 30.23\% and average absolute scenario bias
is 0.05386. These results corroborate the archived calculations while showing
why a pooled coverage number should not be presented as an exact guarantee.
The software paper cites these results as prior evidence and does not count
them as newly rerun simulations.

\subsection{Fixed-score design experiment}

A new fixed-score experiment isolates the inference layer. It crosses target
prevalence 0.5\%, 2\% and 10\%, PSU random-effect standard deviation zero or
0.7, and low or high weight variation. For low variation, each of 20 strata
contains eight PSUs. For high variation, strata alternate between four and
16 PSUs. Each PSU contains 20 population units, two PSUs are sampled without
replacement per stratum, and all units in sampled PSUs are observed.
The inverse inclusion weights and first-stage FPCs are known.

\begin{figure}[t]\centering
\includegraphics[width=.94\textwidth]{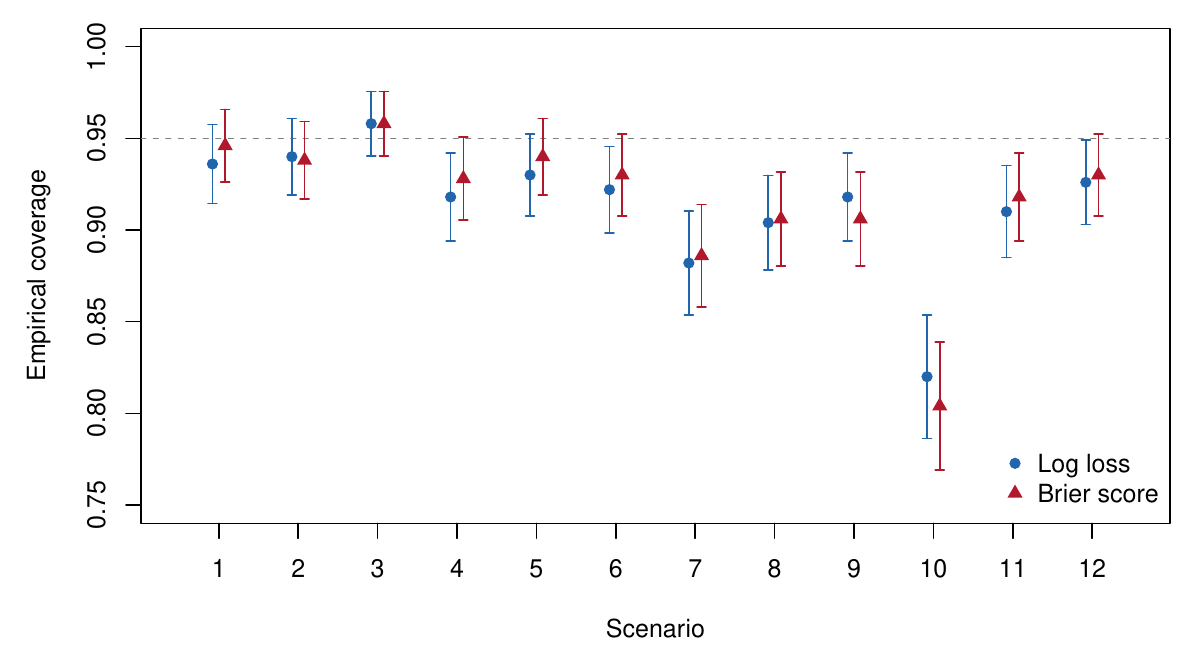}
\caption{Fixed-score coverage in the 12 new sampling scenarios. Error bars represent 1.96 Monte Carlo standard errors. Scenarios 1--6 have low weight variation and 7--12 high variation. Within each block, prevalence increases from 0.5\% to 2\% to 10\%, first with zero PSU effect and then with standard deviation 0.7. The dashed line marks nominal 95\% coverage.}\label{fig:fixed}
\end{figure}

The probability vector is generated before sampling and remains fixed.
Each of the 12 scenarios has 500 independent samples, yielding 6,000 new
sampling replicates. This design tests the score and variance adapters
without confounding them with a learned model. Bias MCSE is the standard
deviation of replicate errors divided by $\sqrt B$; coverage MCSE is
$\{\widehat C(1-\widehat C)/B\}^{1/2}$. At coverage 0.95 and $B=500$,
the latter is about 0.00975. Scenario results appear in
Figure~\ref{fig:fixed} and the replication tables.

Log-loss coverage ranges from 82.0\% to 95.8\%; Brier coverage ranges from 80.4\% to 95.8\%. The lowest values occur with 0.5\% target prevalence, high weight variation and a nonzero PSU effect. Thus, numerical agreement with the reference variance calculation does not ensure near-nominal finite-sample coverage in sparse designs. These intervals use a linearization and Student approximation; the simulation provides a direct check of that approximation.

\subsection{Comparative learning experiment}

The broader software experiment crosses three target prevalences
$(0.005,0.02,0.10)$, two PSU-effect standard deviations $(0,0.7)$,
noninformative versus informative within-PSU selection, one versus three
waves, and target baseline sample sizes 800 versus 2,400. This produces
48 scenarios. Each scenario contains a fixed population of 16,000 subjects,
organized into 20 strata with eight PSUs and 100 subjects per PSU. Three-wave
scenarios contain 48,000 person-wave records. Outcomes depend on two
continuous predictors, an interaction, a quadratic term, a PSU effect,
an optional subject effect and wave. The learning model uses linear
predictor effects, creating deliberate functional-form misspecification.

Sampling selects two PSUs per stratum, followed by Bernoulli selection of
subjects. Informative inclusion probabilities depend on the first predictor
and are normalized to the target expected baseline sample size. Panel
retention depends on the previous outcome and a baseline predictor; a shared
subject-level uniform variate implements monotone dropout. The supplied
analysis weight is the inverse of the exact product of PSU selection,
subject selection and cumulative retention probabilities. This experiment
therefore assesses a known-weight attrition setting, not identification of
unknown response probabilities.

We compare four strategies: unweighted random-fold validation, weighted
random-fold validation, weighted whole-PSU validation, and the fully nested
calibrated workflow. Each uses the same fixed elastic-net specification
$(\alpha=0.5,\lambda=0.01)$ so that differences primarily reflect validation
and calibration choices. This experiment does not compare tuning algorithms
or establish an optimal learner library. The third strategy reuses the raw
outer predictions of the fourth, providing a paired calibration comparison.
Every strategy's final model is evaluated on its complete finite population.

More explicitly, the outcome log odds are
\begin{equation}\label{eq:dgp}
 a+0.7x_1-0.5x_2+0.35x_1x_2+0.2(x_1^2-1)+u_{hc}+b_i+0.1(t-1),
\end{equation}
where $x_1,x_2$ are independent standard normal variables,
$u_{hc}\sim N(0,\sigma_u^2)$, and the panel subject effect is
$b_i\sim N(0,0.4^2)$, with $b_i=0$ in cross-sectional scenarios.
The intercept $a$ is solved numerically so the population mean event
probability equals the target prevalence. Individual-stage inclusion has
probability $\operatorname{logit}^{-1}(a_s+\gamma x_1)$, with
$\gamma=0$ or $0.9$ and $a_s$ solved for the target baseline sample size.
For panels, the conditional retention probability after wave $t$ is
$\operatorname{logit}^{-1}(1.7-0.4x_1-0.5y_{it})$.

Each scenario has 500 repeated samples, for \ComparativeReplicates{} new
sampling replicates and \ComparativeRecords{} strategy records. We retain
every attempted sample and any warning or failure. Metrics include signed
bias, RMSE, conditional score-interval coverage, MCSE, finite-population
calibration intercept and slope, rank performance, fallback counts and
runtime. Because the target is the final full-sample model, CV training-size
effects remain part of the measured discrepancy. The reported conditional
intervals omit training variation; this experiment tests their empirical
behavior rather than assuming unconditional validity.

Across the 48 scenarios, the nested calibrated strategy has mean signed log-loss discrepancy 0.0004, mean scenario RMSE 0.0149, and mean conditional-interval coverage 94.0\%. Its scenario-specific coverage ranges from 84.4\% to 97.8\%. The naive random strategy has mean signed discrepancy 0.0116 and mean coverage 71.3\%.

There are 0 unsuccessful strategy records out of 96,000 attempted records. Paired comparisons of calibrated and raw whole-PSU models are saved for every scenario, with an MCSE for the change in finite-population loss. Calibration lowers mean finite-population log loss in 22 of 48 scenarios; it is not uniformly beneficial.

The lowest nested coverage occurs in scenario 7, with target prevalence 0.5\%, target baseline sample size 800, 1 wave and informative selection. Its coverage MCSE is 1.62 percentage points; the small mean signed discrepancy does not ensure near-nominal interval coverage.

The findings support reporting calibration and uncertainty empirically for the intended design. They do not establish exact unconditional coverage, and the apparent advantages of a procedure should be judged alongside sparse-data fallback counts and scenario-specific errors.

Weighting accounts for most of the average coverage improvement in this
experiment: weighted random and weighted whole-PSU validation average
94.2\% and 94.3\%, respectively. Nested calibration does not dominate either
weighted comparator in mean discrepancy or RMSE. In this fixed linear learner
library, respecting the declared partition target and obtaining a smaller
empirical error are separate considerations. Analysts should choose the
partition to match the intended deployment units and evaluate calibration
within that partition, rather than select a validation scheme because its
observed error is smallest.
\begin{table}[t]
\centering\small
\begin{tabular}{lrrrrr}
\toprule
Strategy & Successful & Bias & RMSE & Coverage (\%) & Scenario range \\
\midrule
Naive random & 24000 & 0.0116 & 0.0220 & 71.3 & 2.6--97.8 \\
Weighted random & 24000 & -0.0001 & 0.0143 & 94.2 & 87.4--97.2 \\
Weighted PSU & 24000 & 0.0003 & 0.0145 & 94.3 & 87.4--97.2 \\
Nested calibrated & 24000 & 0.0004 & 0.0149 & 94.0 & 84.4--97.8 \\
\bottomrule
\end{tabular}
\caption{New comparative experiment. Bias is the mean signed scenario bias in log loss; RMSE and coverage are averaged across scenarios. The range is the minimum and maximum scenario coverage in percent. These conditional intervals do not include full training variation.}\label{tab:comparative}
\end{table}

\begin{figure}[t]\centering
\includegraphics[width=.98\textwidth]{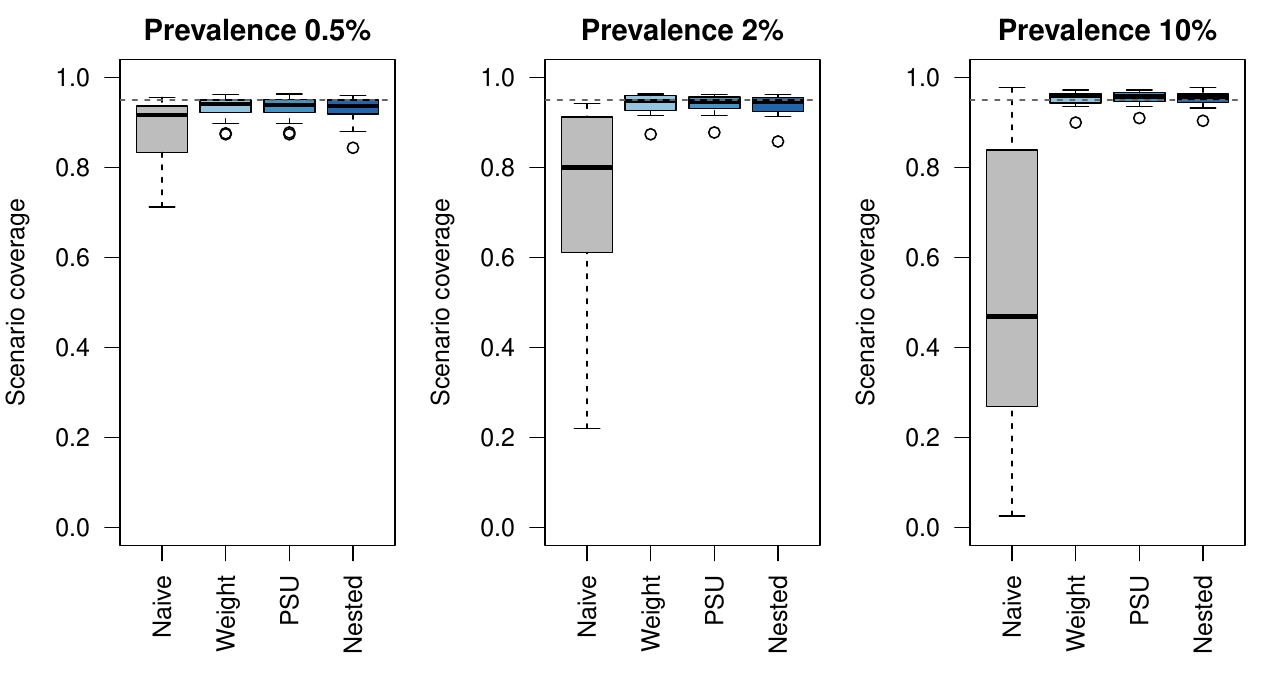}
\caption{Comparative experiment: scenario-specific log-loss coverage by validation strategy and prevalence. The gray line marks nominal 95\% coverage. The distributions summarize scenarios, not independent estimates of a universal coverage parameter.}\label{fig:comparison}
\end{figure}

\subsection{Response-family experiment}

A separate extension experiment crosses continuous, count and three-class
outcomes with one or three waves and noninformative or informative
second-stage sampling, producing 12 scenarios. Each scenario fixes a
population of 4,800 subjects: 20 strata, six PSUs per stratum and 40 subjects
per PSU. Three-wave populations contain 14,400 records. Independent standard
normal predictors $x_1,x_2$, an equiprobable three-level categorical predictor
$g$, PSU effects $u\sim N(0,0.5^2)$ and subject effects
$b\sim N(0,0.3^2)$ generate
\begin{equation}
 \eta_{it}=0.5x_{1i}-0.25x_{2i}+0.2(x_{1i}^2-1)
 +0.2\mathbf{1}\{g_i=B\}+u_i+b_i+0.15(t-1).
\end{equation}
Continuous responses equal $2+\eta_{it}+\epsilon_{it}$ with independent
standard normal errors; counts have conditional mean
$\exp(-0.4+0.5\eta_{it})$. Three-class probabilities are the softmax of
$(\eta_{it},-0.7-0.3x_{1i}+0.25x_{2i},0)$. Ten percent of observed $x_1$
values are missing completely at random. The learner omits the quadratic
and random effects, creating deliberate model misspecification.

Each of 500 samples per scenario selects two PSUs per stratum without
replacement, then samples subjects by independent Bernoulli draws with
probability $\pi_{2i}=0.5$ or
$0.1+0.8\operatorname{expit}(0.8x_{1i}+0.4u_i)$. All waves of selected
subjects are retained, with weight $3/\pi_{2i}$. Weighted inference uses
ultimate-PSU with-replacement linearization and Student intervals; no
census-within-PSU finite-population correction is imposed on this
second-stage Bernoulli sample. The naive comparator uses individual folds,
unit weights and independent-unit variance.

Weighted raw and recalibrated GLMs share the same three-outer/two-inner-fold
fitted models. A separate naive GLM supplies the third learned strategy.
The target is each final model's mean loss over the complete fixed population,
including sampled subjects. A fourth, fixed-score strategy evaluates a
predeclared $x_1$-only predictor, whose definition appears in the replication
script. Those scores are frozen before predictor missingness is imposed;
this diagnostic separates score-variance approximation from model fitting.
Conditional interval coverage and apparent-risk bias are evaluated against
these finite-population targets; outer-score bias is retained separately.
Monte Carlo standard errors use the 500 independent sampling replicates.
The experiment evaluates this specified extension, rather than treating the
binary benchmark as evidence for other response families.
The 6,000 new samples produced 24,000 strategy records, with 0 failed records and 0 records containing warnings. Table~\ref{tab:responses} reports bias, RMSE and conditional interval coverage separately by outcome family.

Recalibration reduced mean finite-population loss in 1 of the 12 scenarios. Paired differences and their Monte Carlo standard errors are retained for every scenario; recalibration is not uniformly beneficial. Across the 12 calibrated scenarios, conditional coverage ranged from 90.0\% to 95.8\%, with maximum estimated coverage Monte Carlo standard error 0.013.

These results caution against assuming that an additional calibration map
improves a fitted mean or probability model. The
\code{calibration = "none"} path remains directly available.
\begin{table}[t]
\centering\small
\begin{tabular}{llrrrr}
\toprule
Outcome & Strategy & Bias & RMSE & Coverage (\%) & Range \\
\midrule
Continuous & Fixed weighted & -0.0006 & 0.0781 & 96.4 & 94.2--97.8 \\
Continuous & Naive random & -0.0223 & 0.0746 & 82.5 & 71.4--95.4 \\
Continuous & Nested raw & -0.0222 & 0.0747 & 92.8 & 90.2--95.8 \\
Continuous & Nested calibrated & -0.0217 & 0.0745 & 93.0 & 90.0--95.8 \\
Count & Fixed weighted & -0.0003 & 0.0372 & 96.2 & 95.8--96.6 \\
Count & Naive random & -0.0024 & 0.0343 & 93.3 & 90.4--95.6 \\
Count & Nested raw & -0.0111 & 0.0339 & 94.3 & 93.8--94.8 \\
Count & Nested calibrated & -0.0111 & 0.0339 & 94.2 & 93.6--94.6 \\
Multiclass & Fixed weighted & 0.0004 & 0.0109 & 96.2 & 94.8--96.8 \\
Multiclass & Naive random & -0.0243 & 0.0291 & 51.4 & 2.4--90.6 \\
Multiclass & Nested raw & -0.0086 & 0.0152 & 93.5 & 90.6--95.6 \\
Multiclass & Nested calibrated & -0.0075 & 0.0134 & 93.7 & 90.6--95.8 \\
\bottomrule
\end{tabular}
\caption{Response-family experiment: four scenarios per family, 500 samples per scenario. Loss is squared error, Poisson deviance or multiclass log loss, respectively; numerical magnitudes are not comparable across families. Bias and RMSE compare sampled mean loss with finite-population risk; learned strategies use apparent predictions. Coverage is for conditional intervals; fixed weighted uses a predeclared predictor. Ranges show scenario coverage in percent.}\label{tab:responses}
\end{table}

\subsection{Evaluation-contract verification and additional sampling experiments}

The extension is checked independently at mathematical boundaries. Exhaustive
enumeration of box-weight vertices validates the sharp sensitivity optimizer;
conformal tests cover the infinity atom, exact cumulative-mass ties and common
weight rescaling. Weighted covariance reconstruction verifies PCA, including
more predictors than observations. Interventions on future measurements leave
trajectory features unchanged. Additional dependency groups are joined
transitively while the original inference identifiers are preserved. Domain
contrasts are checked against the joint covariance, and population counts and
screening ratios against direct \pkg{survey} calculations. Twenty-five
independent \proglang{Python} point-kernel comparisons have maximum absolute
discrepancy \MetricParity{}. Smoother values, domain coefficients and local
explanations are supplied intermediates in this comparison; it does not verify
cross-language equality of their fitting algorithms.

A separate finite-population experiment crosses one versus three waves,
equal versus informative second-stage sampling, and common versus rare outcomes.
Each of eight scenarios uses a fixed population with 24 strata, six PSUs per
stratum and 40 subjects per PSU. Two PSUs are sampled without replacement in
each stratum. Within selected PSUs, simple random samples contain 20 subjects
in the equal-rate condition, or 12 versus 28 according to the prespecified
PSU covariate mean. Inverse two-stage inclusion weights and finite-population
corrections are supplied explicitly. Correlated outcomes arise from PSU and
subject effects; the fixed prediction rule omits these effects. Each scenario
has 1,000 independently seeded samples. Full-population losses, net benefits,
true-positive totals and screening ratios provide the evaluation truths.

Across the 40 metric--scenario combinations, nominal 95\% conditional interval
coverage ranges from \MetricCoverageLow\% to \MetricCoverageHigh\%.
The worst coverage occurs for rare-event net benefit. This is a substantive
finite-sample finding, not a numerical failure: asymptotic Wald intervals
can be inadequate for sparse utility contributions even with the correct
sampling design. Per-scenario Monte Carlo standard errors and all 8,000
sample records are retained.

The conformal experiment separates construction assumptions from empirical
evaluation. Each of 2,000 repetitions per condition has 199 calibration
observations and 100 independent test observations. The fixed prediction is
zero, $Y=X+\varepsilon$, and
$\varepsilon\mid X\sim N\{0,\exp(0.5X)\}$. Calibration covariates are
standard normal. Test covariates are standard normal in the exchangeable
condition, or $N(1,1)$ under shift. The correct shifted density ratio is
$r(x)=\exp(x-1/2)$. A deliberately misspecified construction retains equal
weights under shift. The same shifted datasets are used for the correct and
misspecified weighting conditions.

\begin{table}[tbp]
\centering
\begin{tabular}{lrrr}
\toprule
Construction & Coverage (\%) & MCSE (pp) & Infinite width (\%) \\
\midrule
Exchangeable & 90.09 & 0.082 & 0.000 \\
Known density ratio & 91.43 & 0.117 & 0.494 \\
Shift ignored & 79.47 & 0.105 & 0.000 \\
\bottomrule
\end{tabular}
\caption{Split-conformal experiment, nominal coverage 90\%. Each construction uses 2,000 independent calibration samples and 200,000 test predictions. MCSE is computed across calibration repetitions, accounting for their shared test-set thresholds. Infinite intervals are retained.}\label{tab:conformal}
\end{table}

With nominal 90\% marginal coverage, observed coverage is
\ConformalExchangeable\% for exchangeable data and
\ConformalRatio\% with the known density ratio, versus
\ConformalArbitrary\% when the shift is ignored. The correct weighted method
occasionally returns infinite intervals, as required by its finite-sample
construction. Monte Carlo standard errors use variation across independent
calibration repetitions; they do not treat all test observations sharing a
calibration fit as independent. These results directly motivate the API's
separation of empirical survey weighting from a justified conformal guarantee.

\subsection{Development stress tests and failure accounting}

A separate 600-run development study crosses prevalence, predictor
missingness and one versus three waves. It uses an independent evaluation
sample and deliberately small training data. The initial implementation
failed when an inner \pkg{glmnet} training class had fewer than two
observations. That result motivated the explicit stabilization rule in
Equation~\ref{eq:sparse}. The original failure files are retained in the
replication archive, and all 600 cases are rerun after the change.
The revised implementation completes \StressSuccessful{} of 600 cases.
Numerical completion and statistical adequacy remain separate outcomes:
the audit still reports sparse calibration and engine warnings. This
development experiment is not presented as a prespecified confirmatory
comparison.

\subsection{Sequential-retention validation}

The version 0.3.0 suite has \AttritionTests{} passing test expectations and
\AttritionCoverage\% measured line coverage. Independent weighted logistic
regressions verify the sequential probabilities and cumulative products.
Interventions on held-out outcomes and response histories leave the corresponding
outer training pipeline unchanged. Inner audit records establish disjoint
training groups, and interventions on current/future observations verify lag
eligibility. Tests cover all four response schemas, exact weight preservation
without dropout, stabilization, clipping, invalid roster rejection, serialization
and direct complete-workflow replicate refits.

An additional experiment generates 500 panels under each of four dropout
mechanisms, for 2,000 panels and 6,000 paired strategy records. Each panel contains
640 subjects in 32 equal-size clusters. With independent standard-normal
$X_i,e_{i0},e_{i1},e_{i2}$ and $u_c\sim N(0,0.3^2)$,
\begin{equation}
 Y_{i0}=X_i+u_{c(i)}+e_{i0},\qquad
 Y_{it}=0.7Y_{i,t-1}+0.3X_i+u_{c(i)}+e_{it},\quad t=1,2.
\end{equation}
The baseline weights are $w_{i0}=\exp(0.25X_i)$. Targets are the weighted
complete-panel wave-2 mean and squared-error risk of the fixed predictor $X_i$.
This experiment isolates response selection from a known complete weighted
roster; it does not simulate a new first-stage survey draw or outcome-model fit.
The logit of sequential retention is respectively
\begin{align*}
 1 &\quad\text{(MCAR)},\\
 1-0.65Y_{i,t-1}+0.2X_i &\quad\text{(linear MAR)},\\
 1-0.65Y_{i,t-1}+0.2X_i-0.7(Y_{i,t-1}^2-1)
   &\quad\text{(nonlinear MAR)},\\
 1-0.4Y_{i,t-1}+0.2X_i
 -0.65(Y_{it}-0.7Y_{i,t-1}-0.3X_i) &\quad\text{(MNAR)}.
\end{align*}
The estimated model uses linear lagged $Y$, lagged $X$ and log baseline weight,
four whole-cluster folds, no numerator stabilization and a probability floor of
0.01. Baseline-only complete-case weights and inverse true retention probabilities
are paired comparators. All strategies estimate normalized weighted means; even
oracle weights therefore need not produce an unbiased finite-sample ratio.

\begin{table}[t]
\centering\small
\begin{tabular}{llrrrr}
\toprule
Mechanism & Weights & Mean bias & MCSE & Mean RMSE & Loss RMSE\\
\midrule
MCAR & Baseline only & -0.0019 & 0.0031 & 0.0699 & 0.1153\\
MCAR & Oracle & -0.0019 & 0.0031 & 0.0699 & 0.1153\\
MCAR & Estimated & 0.0003 & 0.0021 & 0.0477 & 0.1179\\
Linear MAR & Baseline only & -0.5666 & 0.0035 & 0.5719 & 0.1531\\
Linear MAR & Oracle & -0.0137 & 0.0064 & 0.1446 & 0.2508\\
Linear MAR & Estimated & 0.0262 & 0.0063 & 0.1434 & 0.2921\\
Nonlinear MAR & Baseline only & -0.4248 & 0.0044 & 0.4358 & 0.4084\\
Nonlinear MAR & Oracle & -0.2150 & 0.0166 & 0.4282 & 0.6451\\
Nonlinear MAR & Estimated & -0.2981 & 0.0037 & 0.3093 & 0.4385\\
MNAR & Baseline only & -0.6350 & 0.0035 & 0.6397 & 0.1689\\
MNAR & Oracle & 0.0016 & 0.0066 & 0.1483 & 0.3508\\
MNAR & Estimated & -0.3286 & 0.0030 & 0.3353 & 0.2390\\
\bottomrule
\end{tabular}
\caption{Sequential-retention experiment: 500 panels per mechanism, with paired weighting comparisons. MCSE refers to mean-estimation bias. Loss is squared error of the fixed predictor $X$. All 6,000 strategy records are retained.}\label{tab:attrition}
\end{table}

Under linear MAR, estimated weighting reduces absolute mean bias from 0.5666
to 0.0262 (MCSE 0.0063). Mean-estimation RMSE falls from 0.5719 to 0.1434, while
fixed-predictor loss RMSE increases from 0.1531 to 0.2921. The nonlinear mechanism
both misspecifies the fitted model and can yield very small retention probabilities;
even the oracle ratio has appreciable finite-sample bias and variability.
Under MNAR, estimated mean bias remains $-0.3286$ (MCSE 0.0030).
These paired results show the benefit under the specified observed-history model
without treating weighting as uniformly beneficial across metrics or as a remedy
for dependence on unobserved current outcomes. Full replicate records, bias MCSEs,
RMSEs, effective sample sizes, clipping counts and exact simulation sources are
retained with the replication files.

\section{Computational performance and reproducibility}\label{sec:performance}

All local measurements use \proglang{R} 4.4.2 on Windows 11 with 24 logical
processors and approximately 64 GiB physical memory. Optional package
versions, input hashes, session information and commands are recorded in
the replication archive. Parallel comparative simulations use eight
independent worker processes with deterministic scenario and replicate
seeds. Checkpoints are written in batches; failed runs remain identifiable.
The original workload measurements below describe version 0.1.0. The version
0.2.0 extension is separately measured in 25 fresh processes, with median
recorded attachment time \MetricLoadMedian{} seconds (range
\MetricLoadMin{}--\MetricLoadMax{} seconds); all processes verify that optional
survey and modeling namespaces remain unloaded. Recorded zero times reflect
timer resolution, not instantaneous loading. The original timing records
are retained rather than silently reassigned to a changed implementation.

Twenty-five fresh-process measurements give a median package attachment
time of \LoadMedian{} seconds and a range of \LoadMinimum{}--\LoadMaximum{}
seconds. Each process asserts that attachment does not load \pkg{survey},
\pkg{Matrix}, \pkg{glmnet}, \pkg{ranger}, \pkg{xgboost}, \pkg{nnet} or
\pkg{reticulate}. These are fresh \proglang{R} process measurements with the
operating-system file cache uncontrolled; they are not cold-disk timings.
Short elapsed times are subject to the operating system's timer resolution.

The score kernel processes one million binary predictions and weights in
\MillionMetric{} seconds at the median of five repetitions. Its input vectors
occupy approximately 20 MB. Weighted ROC and average-precision computations
sort scores and aggregate ties, requiring $O(n\log n)$ time in the comparison
sorting model and $O(n)$ storage. Proper scores alone require linear work.
This benchmark evaluates metrics, not a million-record model fit.
Separate one-million-record score measurements give medians of
\MillionGaussian{} seconds for continuous responses, \MillionPoisson{}
seconds for counts and \MillionMultinomial{} seconds for three-class
probabilities, each over five repetitions. These family-specific kernels
perform linear work and omit the binary rank-metric sorting operation.

An additional stress benchmark learns and applies the weighted numeric
recipe to one million rows and 500 predictors. The input and output each
occupy approximately 4.0 GB. Recipe estimation takes \LargeFit{} seconds
and application takes \LargeBake{} seconds. The R process reports a peak
working set of \LargePeak{} GiB. Dense encoding therefore
has a real $O(np)$ storage cost. Table~\ref{tab:speed} presents measured
workloads without extrapolating them to every optional learner.

\begin{table}[t]
\centering\small
\begin{tabular}{lrrrr}
\toprule
Operation & Rows & Predictors & Seconds & Input MB \\
\midrule
Binary metrics &     1,000 & 1 & 0.00 & 0.02 \\
Binary metrics &    10,000 & 1 & 0.00 & 0.20 \\
Binary metrics &   100,000 & 1 & 0.05 & 2.00 \\
Binary metrics & 1,000,000 & 1 & 0.39 & 20.00 \\
Recipe fit and apply &     1,000 & 10 & 0.00 & 0.08 \\
Recipe fit and apply &     1,000 & 100 & 0.02 & 0.81 \\
Recipe fit and apply &     1,000 & 500 & 0.09 & 4.06 \\
Recipe fit and apply &    10,000 & 10 & 0.01 & 0.80 \\
Recipe fit and apply &    10,000 & 100 & 0.11 & 8.01 \\
Recipe fit and apply &    10,000 & 500 & 0.66 & 40.06 \\
Recipe fit & 1,000,000 & 500 & 67.89 & 4000.06 \\
Recipe apply & 1,000,000 & 500 & 11.59 & 4000.06 \\
\bottomrule
\end{tabular}
\caption{Measured computational workloads. Binary metric timings are medians of five repeats; recipe workloads are individual stress measurements. Zeroes reflect timer resolution. Input size is object storage, not peak process memory. Large recipe timings were measured alongside the simulation workload.}\label{tab:speed}
\end{table}

For version 0.3.0, another 25 fresh processes give median recorded attachment
time 0.01 seconds (range 0.00--0.04), again without loading optional engines.
The standalone sequential-retention helper processes 999,999 scheduled records
from 333,333 subjects in median 13.89 seconds over three runs (range
13.31--14.44). This includes roster validation and two weighted logistic fits on
a preconstructed design; it does not include nested outcome learning. These
measurements retain the same machine and timer qualifications as the earlier
benchmarks.

The source package contains documentation, examples and tests, while raw
research inputs, simulation records and the software paper remain outside
the installed package. The reproducibility scripts identify which outputs
come from the prior publication and which are newly generated. They verify
source hashes before constructing the NHANES cohort and write new results
to separate locations. Independent \proglang{Python} code supplies reference
score and first-stage variance calculations. Cross-platform automated-check
configuration is included; the release record links the hosted check results
and distinguishes them from locally observed checks.

\newpage
\section{Discussion}\label{sec:discussion}
\pkg{svylearn} makes survey prediction workflows inspectable at the boundaries
where silent errors commonly arise: sampling-unit partitioning, training-fold
preprocessing, model selection, recalibration and design-based inference. Its
primary software advance over ad-hoc scripts is an explicit, testable contract
governing how these operations compose. It orchestrates established engines
(\pkg{survey}, \pkg{glmnet}, \pkg{ranger}, \pkg{xgboost}, \pkg{nnet}) behind a
unified S3 interface that enforces declared partition boundaries and records
sparse-data stabilization rules. The published method remains the methodological
foundation; the new software experiments provide separately attributable evidence.

\subsection{Applicability across survey and weighted data ecosystems}

Because design metadata are separate from domain-specific variable schemas,
the interfaces accommodate several data architectures. The following examples
describe interface compatibility; NHANES and ECLS-K:2011 are the completed
real-data applications in this paper. Each dataset still requires the variables,
weights and replication conventions appropriate to its particular release.

\begin{description}
\item[Multistage household and health surveys.]
For datasets such as NHANES, DHS and NSDUH, declaring highest-stage PSUs and
strata in \code{svy_design()} keeps entire declared clusters intact across
outer and inner partitions. Additional dependence groups can join related
units when the released sampling identifiers alone do not represent all
relevant household or spatial dependence.
\item[Surveillance and replicate-weight designs.]
For architectures encountered in BRFSS, MEPS and ACS, analysts can supply
the appropriate Taylor variables or agency replicate matrices and conventions,
including BRR, Fay, jackknife and bootstrap constructions. Raked margins and
externally known event totals are audited separately by \code{svy_rake_audit()},
whose output can enter \code{svy_eval()}.
\item[Longitudinal cohorts and panels.]
Subject and wave identifiers keep repeated records together in structures such
as ECLS-K, HRS and PSID. Historical summaries use \code{svy_trajectory()} with
a declared time cutoff. Analysts can use agency longitudinal weights, as in
our ECLS application, or estimate sequential retention weights from a complete
baseline roster using \code{svy_attrition_weights()}. Passing an unfitted
\code{svy_attrition()} specification to \code{svy_fit()} refits the retention
model inside each inner and outer training partition. These are alternative
weighting routes; an agency's existing attrition adjustment is not applied twice.
\item[Weighted real-world evidence and transport diagnostics.]
Unclustered weighted designs use \code{ids = NULL}. Supplied source/target
weights can be checked with covariate SMDs and prediction divergence, and
\code{svy_weight_bounds()} explores bounded perturbations of weighted utilities.
Inverse-treatment or estimated-selection weights in EHR or claims data do not
become probability-sampling weights simply through this interface. Causal
identification and nuisance-weight variance require their own analysis; the
package does not turn a prediction audit into a causal effect estimator.
\end{description}

\subsection{Methodological boundaries and interpretive contracts}

Whole-PSU partitioning addresses declared sampling dependence, while weighting
and sequential retention modeling address selection under stated assumptions.
The new attrition component actively estimates observable time-varying retention
under sequential missing at random, with automatic lag construction and fold
isolation. Its identification still requires adequate recorded history,
positivity and an appropriate conditional response model. The simulation
includes misspecification and MNAR conditions to assess what the fitted weights
actually achieve. Bounded-weight utility perturbations are a different
sensitivity target from an explicit non-ignorable dropout model
\citep{Su2022}.

Recalibration is an estimated mapping with its own support requirements. It can
improve proper scores and probability agreement, leave discrimination unchanged
or slightly degraded, or require intercept-only fallback under extreme rarity.
Conditional fixed-score standard errors quantify sampling uncertainty for given
predictions and weights. They omit model selection, outcome refitting and
retention-model estimation. With suitable baseline-roster replicate weights,
\code{svy_refit_replicates()} reruns the complete nested statistic, including
retention models, at greater computational cost. Its validity remains tied to
the supplied replication scheme and target. Within-PSU permutation importance
and aggregated SHAP summaries quantify associational model reliance.

\subsection{Scope and further development}

The current release supports binary, continuous, count and nominal multiclass
outcomes under cross-sectional and panel designs, including estimated monotone
retention weighting. Right-censored time-to-event outcomes and competing risks
require dedicated IPCW proper scores and calibration diagnostics. Native pooling
across multiply imputed surveys, nonmonotone observation models, automatic
transport-weight estimation and smooth weight-trimming sensitivity envelopes
are separately specified extensions. Each has an estimand, assumptions and
validation requirements distinct from the completed sequential-retention work.

\subsection{Conclusion}

The lightweight architecture, recorded failure diagnostics, explicit estimands
and scenario-specific benchmarks provide an auditable basis for adoption.
Analysts can use individual components---including \code{svy_eval()},
\code{svy_attrition_weights()}, \code{svy_conformal()} and
\code{svy_sensitivity()}---or deploy the full nested workflow. The retained
numerical checks and simulation outcomes make its behavior reviewable across
the supported designs and response families.

\section*{Computational details}
The original experiments use \pkg{svylearn} 0.1.0. The evaluation-contract and
ECLS-K:2011 additions use the separately recorded 0.2.0 implementation and
frozen metric-simulation runtime. The sequential-retention addition is version
0.3.0, with separately recorded tests and dropout experiments. Software versions are in the
machine-readable session manifest. Replication commands are supplied in
\code{scripts/}; all paper tables and figures are generated from saved result
files. The method citation in \code{inst/CITATION} follows the title,
authors, issue and DOI printed in the supplied published article.

\begingroup
\interlinepenalty=10000
\bibliography{references}
\endgroup
\end{document}